\documentclass{elsart}
\usepackage{natbib,epsfig}
\usepackage{amsfonts}

\begin{document}
\runauthor{Salzmann and Ralchenko}
\begin{frontmatter}
\title{New parametrization for differences between plasma kinetic codes}
\author{David Salzmann\thanksref{aff}} and
\author{Yuri Ralchenko\thanksref{corresp_author}}
\address{Atomic Physics Division, National Institute of Standards and Technology,
Gaithersburg, MD 20899-8422}
\thanks[aff]{\emph{Permanent address}: Soreq Nuclear Research Center, Yavne,
Israel 81800.}
\thanks[corresp_author]{\emph{Corresponding author. Email}:
yuri.ralchenko@nist.gov.}

\begin{abstract}
Validation and verification of plasma kinetics codes 
requires the development of quantitative methods and techniques for code
comparisons. We describe two parameters that can be
used for characterization of differences between such codes. It is shown
that these parameters, which are determined
from the most general results of kinetic codes, can provide
important information on the differences between the basic rate coefficients
employed. Application
of this method is illustrated by comparisons of some results from the 3rd
NLTE Code Comparison Workshop for carbon, germanium, and gold plasmas.
\end{abstract}



\end{frontmatter}

\section{\label{sec:intro}Introduction}

Spectroscopy of plasmas has a wide range of applications \cite{book,Griem},
including the study of astrophysical objects, diagnostics of laser produced
plasmas, and analysis of radiation emission from plumes of rockets. In spite of
the importance of this subject and the substantial efforts made by numerous
groups, there persist significant discrepancies in results of the plasma kinetic
codes used to analyze plasma emission spectra under
non-local-thermodynamic-equilibrium (NLTE) conditions. For examples of the level
of disagreement see, e.g., the reports from the NLTE Code Comparison Workshops
\cite{NLTE1,NLTE2,NLTE3}.

Given the plasma particle temperature and density, the first task of kinetic
codes is the computation of the charge state and excitation state distributions
\cite{book,Griem}. These are typically determined from the set of rate
equations:

\begin{eqnarray}
\frac{d\,N_{\zeta m}}{dt} & = &\sum_{all ~ populating ~ processes}
{n_{e}^{k}N_{\zeta ^{\prime }m^{\prime }}\mathbb{R}_{\zeta
^{\prime }m^{\prime }\rightarrow \zeta m}} - \nonumber \\
& & \sum_{all ~ depopulating ~ processes}{n_{e}^{k}N_{\zeta m}\mathbb{R}_{\zeta
m\rightarrow \zeta ^{\prime }m^{\prime }}},  \label{Eq1}
\end{eqnarray}
where $N_{\zeta m}$ is the density of ions having charge $\zeta $ ( $%
0\leq \zeta \leq Z$ with $Z$ being the nuclear charge) 
excited to state $m$ (ordered according to their
ascending energy, $m=0$ corresponds to the ion's ground state), $n_{e}$ is
the electron density, and $\mathbb{R}_{a\rightarrow b}$ denotes 
the rate coefficient (r-c) for the transition of an ion from state $a$ to
state $b$ due to some atomic process. In Eq. (\ref{Eq1}), $k$ represents the
number of
electrons taking part in any given process, thus, $k=0$ for spontaneous
decay and autoionization, $k=1$
for the electron impact processes (e.g., excitation and photorecombination), and
$k=2$ for the three-body recombination. In large-size plasmas, where linear
dimension is larger than the photon mean free path, photon induced
processes must also be included in (\ref{Eq1}). In a steady state plasma,
$d\,N_{\zeta m}/dt=0$, and Eq. (\ref{Eq1}) reduces to:

\begin{equation}
\sum_{all ~ populating ~ processes}{
n_{e}^{k}\,N_{\zeta ^{\prime }m^{\prime }} \mathbb{R}_{\zeta ^{\prime
}m^{\prime }\rightarrow \zeta m}} = \sum_{all  ~ 
depopulating  ~ processes}{n_{e}^{k}N_{\zeta m} \mathbb{R}_{\zeta
m\rightarrow \zeta ^{\prime }m^{\prime }}}.  \label{Eq2}
\end{equation}
This is a finite set of non-linear coupled equations for $N_{\zeta m}$
's. If one is interested only in the density of the charge states,
regardless of the ionic excitations, Eq. (\ref{Eq2}) can be further simplified, and
the result is a set of recursive equations:

\begin{equation}
\frac{N_{\zeta +1}}{N_{\zeta }}=\frac{I_{\zeta \rightarrow \zeta +1}}{%
R_{\zeta +1\rightarrow \zeta }^{(2)}\,+\,\,\,n_{e}\,R_{\zeta +1\rightarrow
\zeta }^{(3)}}.  \label{Eq3}
\end{equation}
Here $N_{\zeta }=\sum_{m}N_{\zeta m}$ is the density of the charge
state, and \ $I_{\zeta \rightarrow \zeta +1}$, $R_{\zeta +1\rightarrow \zeta
}^{(2)}$ and $R_{\zeta +1\rightarrow \zeta }^{(3)}$ are the \textit{total
rate coefficients} for ionization, two-body (radiative+dielectronic)
recombination, and three-body recombination, respectively. For quasineutral
plasmas, solutions of
Eq.(3) have to satisfy two complementary conditions, namely,

\begin{equation}
n_{i}=\sum\limits_{\zeta =0}^{Z}\,N_{\zeta }\;;\quad
n_{e}=\sum\limits_{\zeta =0}^{Z}\,\zeta \,N_{\zeta }=\overline{Z}\,n_{i}.
\end{equation}
Here $n_i$ is the total ion density, and $\overline{Z}$ is the
average charge of the ions. In the following we assume that $N_{\zeta}$
is the fractional abundance of charge state $\zeta$, which is equivalent to
the assumption that $n_i$ = 1.

The disagreement between the results of the kinetic codes developed by various
researchers may originate from several factors. One of the most important is the
approximate character of the r-c's. In the literature one can find several
recommended formulas for the relevant r-c's. Some may be more accurate than
others, but none was shown to have high accuracy. For higher accuracy, the r-c's
can be directly calculated for each transition between atomic states, using
advanced quantum-mechanical methods. The necessity to generate a large number of
r-c's for kinetic calculations, however, impedes the application of these
techniques, thereby forcing a compromise between computational speed and
accuracy. The solutions of (\ref{Eq3}) obviously depend on the methods chosen
for the determination of the r-c's in the right-hand side of (\ref{Eq3}).

Another source of disagreement is the criterion used for the continuum or ionization
potential lowering in plasmas. This is particularly important in high-density
plasmas where the plasma potential moves the upper ionic bound states into the
continuum, leaving the ion only with a finite number of discrete states.
Moreover, due to the fluctuations of the local microfield around each ion, even
the "tightly bound states" may change into instantaneous quasi-molecular states
whose treatment is, as yet, not clear.

\begin{figure*}[t]
\begin{minipage}{15cm}
\epsfig{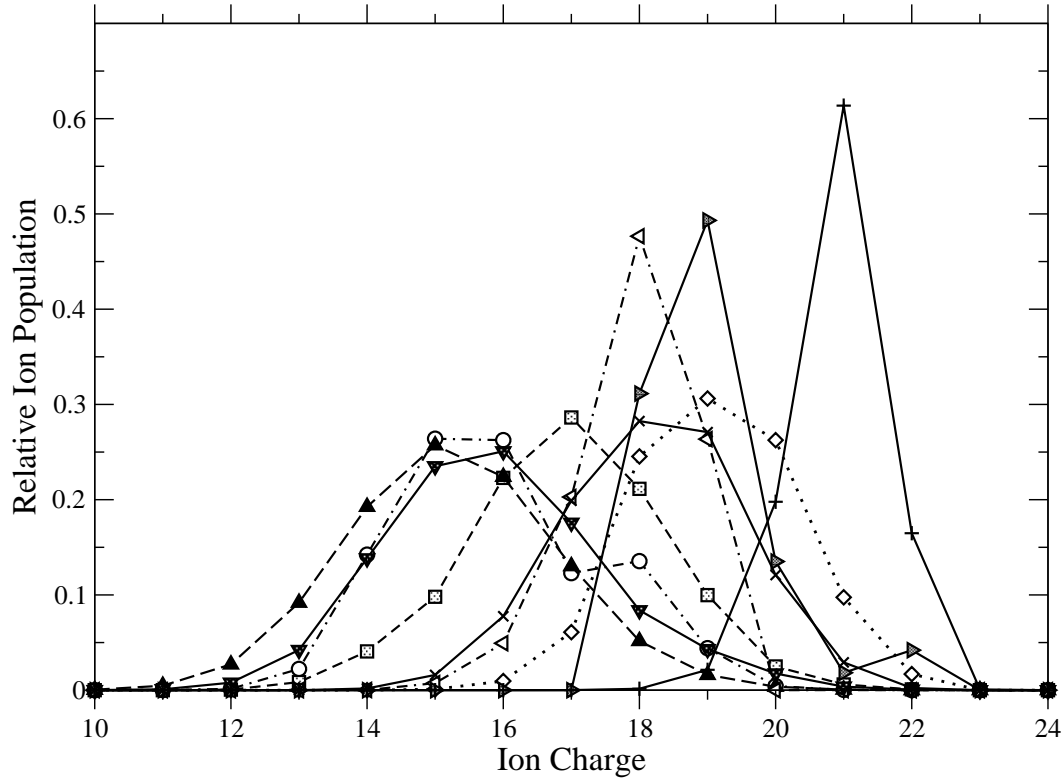}
\end{minipage}
\vspace*{1cm}
\caption{Relative ion populations for a steady-state germanium plasma at electron
temperature T$_e$ = 250 eV and electron density N$_e$ = $10^{17}$ cm$^{-3}$
\cite{NLTE3,SAHA}.}
\label{fig1}
\end{figure*}

Comparison of the results of NLTE kinetic codes was a subject of several
workshops \cite{NLTE1,NLTE2,NLTE3}. The participants were asked to submit large
sets of various physical quantities to be compared, and in numerous cases very
significant differences were found. This situation is well exemplified in Fig.
\ref{fig1}, where the various calculations of the relative ion populations for a
germanium plasma are presented for a specific case from the NLTE-3 Workshop
\cite{NLTE3}. One can clearly notice a significant spread  both for the mean ion
charges and for the distribution widths calculated with different kinetic codes.
Although a variety of physical parameters investigated at the Workshop was
mostly sufficient to draw conclusions about sources of discrepancies,
introduction of simple and clearly defined new parameters that would pinpoint
some fundamental underlying differences would greatly facilitate such
comparisons. 


In Ref. \cite{PRA} one of the authors developed a method that allows 
determination of variations in an ionization state distribution due to small
changes in the rates of basic atomic processes. The purpose of the present paper
is the opposite, that is, to introduce new parameters that quantify differences
in the {\it input} atomic rates in kinetic codes using the {\it results} of
calculations. Such approach is similar to solution of the inverse problem in
physics. 

Below we address a simple but critical question, namely: ``How different are two
population kinetic codes?" It is obvious that the most comprehensive answer can
be obtained provided all input and output parameters as well as the complete
description of the approximations used are available. Unfortunately, this is not
always the case. Here we introduce parameters that are straightforward to
calculate and can provide an answer in a clear manner. An important feature of
the proposed method is that only the most general kinetic characteristics, such
as the mean ion charge and central momenta, are required to determine these new
parameters.

The paper is organized as follows. In the next section we define these
parameters and explain their meanings. In Section \ref{sec3} we apply the method
to characterize the differences between code results for three representative
cases from the NLTE-3 Workshop \cite{NLTE3}. Finally, in Section \ref{sec4}, a
short summary and conclusions are presented.

\section{\label{sec2}Quantitative characterization of kinetic codes}

We introduce the notation $Q_{\zeta }$ for the
r.h.s. of Eq.~(\ref{Eq3}):

\begin{equation}
Q_{\zeta }=\frac{I_{\zeta \rightarrow \zeta +1}}{R_{\zeta +1 \rightarrow
\zeta }^{(2)} + n_e R_{\zeta +1\rightarrow \zeta }^{(3)}}
\end{equation}
and denote by $\Delta Q_{\zeta } \equiv Q_{ref} - Q$ the overall difference in
this quantity for 
a given code relative to the reference. 
Using this definition, Eq.(3) is rewritten as:

\begin{equation}
\frac{N_{\zeta +1}}{N_{\zeta }}=Q_{\zeta }.
\end{equation}
It was shown in Ref. \cite{PRA}, that the relative difference in the final
results, $\Delta N_{\zeta }/N_{\zeta }$, caused by the relative differences
used in the input data, $\Delta Q_{\zeta }/Q_{\zeta }$, is given by

\begin{equation}
\frac{\Delta N_{\zeta }}{N_{\zeta }}{\ =\alpha }_{\zeta }\,{\ -\,}\overline{
\alpha },  \label{dn1}
\end{equation}
where

\begin{equation}
{\alpha }_{\zeta }=\sum\limits_{\zeta ^{\prime }=0}^{\zeta }\frac{\Delta
Q_{\zeta ^{\prime }}}{Q_{\zeta ^{\prime }}}\;; ~~ \\
\overline{\alpha }=\sum\limits_{\zeta =0}^{Z}\alpha _{\zeta }\frac{%
N_{\zeta }}{n_{i}}. \label{al1}
\end{equation}
In particular, if the relative differences are all equal, then
\begin{equation}
\Delta Q_{\zeta}/Q_{\zeta } \equiv p=const, \label{dQ}
\end{equation}
and Eq.(\ref{dn1}) reduces to the simple form \cite{PRA}:

\begin{equation}
\frac{\Delta N_{\zeta }}{N_{\zeta }} = p(\zeta -\overline{Z}).  \label{dn0}
\end{equation}
In Ref. \cite{PRA} a full discussion is presented about the meaning of
Eq. (\ref{dn0}), as well as how and when this difference influences the
results of kinetic codes. Even if $\Delta Q_{\zeta }\bigskip /Q_{\zeta }$ is
not constant, one can define their average value, $p\equiv \left\langle
\Delta Q_{\zeta }\bigskip /Q_{\zeta }\right\rangle _{\zeta }$, over the
relevant ion charge states.

The above formulas were derived assuming small differences between
code results. In a general case, however, the deviations are not small, and one
has to symmetrize the relevant parameters with respect to both compared codes, i.e.:

\begin{equation}
N_{\zeta} = \frac{1}{2}(N_{1,\zeta}+N_{2,\zeta}), 
~ \overline{Z} = \frac{1}{2} (\overline{Z}_1+\overline{Z}_2),
\end{equation}
and so on, which is assumed in what follows.

Multiplying both sides of Eq.(\ref{dn0}) by $(\zeta -\overline{Z}) N_{\zeta}$ and summing over
all charge states, one arrives at:
\begin{equation}
p = \frac{\Delta \overline{Z}}{\overline{Z^2} - {\overline{Z}}^2} \equiv
\frac{\Delta \overline{Z}}{\sigma_2} \label{newp}
\end{equation}
where $\sigma_2$ is the variance, or second central moment, which is related to
the ionization distribution width. Remarkably, this equation 
links the average difference in the ratios of effective 
atomic rates to the most general plasma parameters, namely, mean ion charges
 and variances.

However, if two compared codes have the same $\overline{Z}$ but different
variances, the parameter $p$ defined by Eq. (\ref{newp}) is zero. This simply
means that the approximation (\ref{dQ}) is insensitive to differences in
ionization distribution widths. To take this dependence into account, one can
add the next term in the expansion with respect to $(\zeta - \overline{Z})$:

\begin{equation}
\frac{\Delta Q_{\zeta }}{Q_{\zeta }}=p+2k(\zeta -\overline{Z}). \label{dQ1}
\end{equation}
Substituting (\ref{dQ1}) into Eqs. (\ref{al1}) and (\ref{dn1}), one obtains:

\begin{equation}
\frac{\Delta N_{\zeta }}{N_{\zeta }} = p(\zeta - \overline{Z}) + k
\left[(\zeta - \overline{Z})^2 + (\zeta - \overline{Z}) - \sigma_2\right].  \label{dnn}
\end{equation}
To derive equations relating two parameters, $p$ and $k$, we multiply Eq.
(\ref{dnn}) by $(\zeta -\overline{Z}) N_{\zeta}$ and $(\zeta -\overline{Z})^2
N_{\zeta}$ and then sum both sides over ion charges. The two ensuing equations are
sufficient to
obtain the following expressions:

\begin{eqnarray}
p &=& \frac{\Delta \overline{Z} \left(\sigma_4 + \sigma_3 - \sigma_2^3\right) 
- \Delta \sigma_2 \left(\sigma_2 + \sigma_3\right)}{\sigma_4 \sigma_2 - \sigma_3^2 -
\sigma_2^3},  \nonumber \\
k &=& \frac{\Delta \sigma_2 \cdot \sigma_2 - \Delta \overline{Z} \cdot \sigma_3}
{\sigma_4 \sigma_2 - \sigma_3^2 - \sigma_2^3}, \label{newkp}
\end{eqnarray}
where 
$\sigma_i = \sum_{\zeta =0}^{Z_{max}} \left(\zeta - \overline{Z}\right)^i N_{\zeta}
$
is the $i$th central moment.

Equation (\ref{newkp}) is the main result of the present paper. We propose to
use the parameters $p$ and $k$ for characterization of differences between
plasma kinetic codes. Using Eq. (\ref{newkp}), which depends only on the most
general kinetic parameters, namely, mean ion charges and central momenta, one
can directly evaluate the average differences between the effective rates
implemented in various kinetic codes.

\section{\label{sec3}Comparison of kinetic codes}

The above described method is applied here to the computational results from the
3rd NLTE Workshop \cite{NLTE3} that can be accessed in the NIST SAHA database
\cite{SAHA}. This database contains various parameters, including mean ion
charges, central momenta, and ion populations, which may be used to determine
the quantities $p$ and $k$ defined in Eqs. (\ref{newp}) and (\ref{newkp}). The
SAHA database also provides other valuable parameters, such as effective rates
and partition functions, so that a user can obtain a deep insight into
differences between kinetic codes. In accordance with the policy accepted by the
Workshop participants, the results will be presented without direct attribution,
although a list of participating codes will be given for each case. Note also
that not all codes provide a complete set of central momenta up to $\sigma_4$.

The first step of the comparison procedure consists in selection of a reference
against which the other codes are to be compared. In the following
comparisons the reference code is chosen arbitrarily, as generally there are no
{\it a priori} physical grounds to prefer a particular code. Obviously, the
average values of $p$ and $k$ would change when selecting another code as a
reference. However, the standard deviations $\sigma$ of the corresponding
distributions of $p$ and $k$ that reflect the average spread within a group of
codes should not change and therefore are reported below as well. In what
follows, the parameters $p$ and $k$ determined from Eqs. (\ref{newp}) and
(\ref{newkp}) are referred to as ``calculated", while those determined by
fitting the $\Delta Q/Q$ ratios (Eqs. \ref{dQ} and \ref{dQ1}) are referred to as
``fit". As the fitting procedure has to include only the physically
significant cases, ion states with $N_\zeta < 10^{-4}$ were excluded from the
comparison. Also, in order to emphasize the contribution from the most populated
states, the fitting was performed with the weights $g_w = \sqrt{N_1 N_2}$, where
$N_1$ and $N_2$ are the ion populations of the reference and compared codes.

Among numerous cases available in the SAHA database we selected three cases for
germanium, carbon, and gold. These elements cover a wide range of ion
charges and their ions represent atomic systems with different level of
complexity.

\subsection{Ge}

For germanium, we selected a relatively simple case of T$_e$ = 600 eV and N$_e$
= $10^{17}$ cm$^{-3}$, where almost all codes have a mean ion charge $\overline
Z \approx 22$ corresponding to a closed-shell Ne-like Ge. This case will be
discussed in more detail than the C and Au cases.

Table \ref{TbGe} presents calculated and fit (superscript ``$f$") values of
parameters $p$ and $k$. The values of $p$ determined from the single-parameter
formula (\ref{newp}) or fit using Eq. (\ref{dQ}) have subscript ``0". One can
immediately notice a generally good agreement between the calculated and fit
values of $p$ for all but one code. Table  \ref{TbGe} clearly demonstrates that
code 7 is an outlier, which is also emphasized by its very different value of
the mean ion charge $\overline Z \approx 27$. Moreover, the two-parameter fit
was not performed for code 7, as for only one ion stage both code 7 and the
reference code have populations larger than $10^{-4}$.

\begin{table}
\caption{Calculated and fit parameters $p$ and $k$ for the Ge case of T$_e$ = 600
eV and N$_e$ = $10^{17}$ cm$^{-3}$. Superscript $f$ denotes fit values.
Subscript 0 denotes $p$'s determined from the one-parameter formulas (\ref{dQ})
and (\ref{newp}). Code 7 was excluded in determination of standard deviation
$\sigma$.}
\label{TbGe}
\begin{center}
\begin{tabular}{c|cc|cccc}
\hline 
Code No. & $p_0$ & $p_0^f$ & $p$ & $k$ & $p^f$ &$k^f$ \\
\hline 
1 & -0.566 & -0.708 & -0.509 & -0.249 & -0.510 & -0.308 \\
2 & -0.400 & -0.247 & -0.379 & -0.044 & -0.309 &  0.124 \\
3 & -0.507 & -0.417 & -0.571 &  0.130 & -0.594 &  0.288 \\
4 &  0.223 &  0.222 &  0.159 &  0.150 &  0.140 &  0.256 \\
5 & -0.203 & -0.140 & -0.210 &  0.033 & -0.206 &  0.113 \\
6 & -0.764 & -0.942 & -0.760 & -0.130 & -0.760 & -0.260 \\
7 & -4.266 & -1.298 & -3.800 & -1.389 &  ----- &  ----- \\
8 & -0.592 & -0.122 & -0.334 & -0.175 & -0.018 &  0.413 \\
\hline 
$\sigma$ & 0.301 & 0.362 & 0.272 & 0.142 & 0.298 & 0.255 \\
\hline 
\end{tabular}
\end{center}
\end{table}

Another interesting feature is a very small difference between the ``simple" $p$
of Eq. (\ref{dQ}) and calculated and fit values of $p$ determined from the
two-parameter formulas. While agreement between differently calculated
parameters $p$ is generally very good, the calculated and fit values of $k$
show worse level of correspondence. This is not surprising since $k$ is
a high-order parameter which may be more sensitive to small variations in
data.

As already mentioned, a standard deviation $\sigma$ would unambiguously
represent the spread of parameters $p$ and $k$ within a particular group of
codes. The last row in Table \ref{TbGe} show $\sigma$'s calculated for each
column (the outlier code 7 was not included in the determination of $\sigma$).
Remarkably, $\sigma$'s for calculated $p_0$ and two-parameter calculated and fit
$p$'s agree within only 6 \%, and even $\sigma(p_0^f)$ deviates by less than 25
\%. The value of $\sigma \approx $ 0.3 means that in this group of codes the
average deviation of effective ionization and recombination rates is about 30 \%.

\begin{figure*}[t]
\epsfig{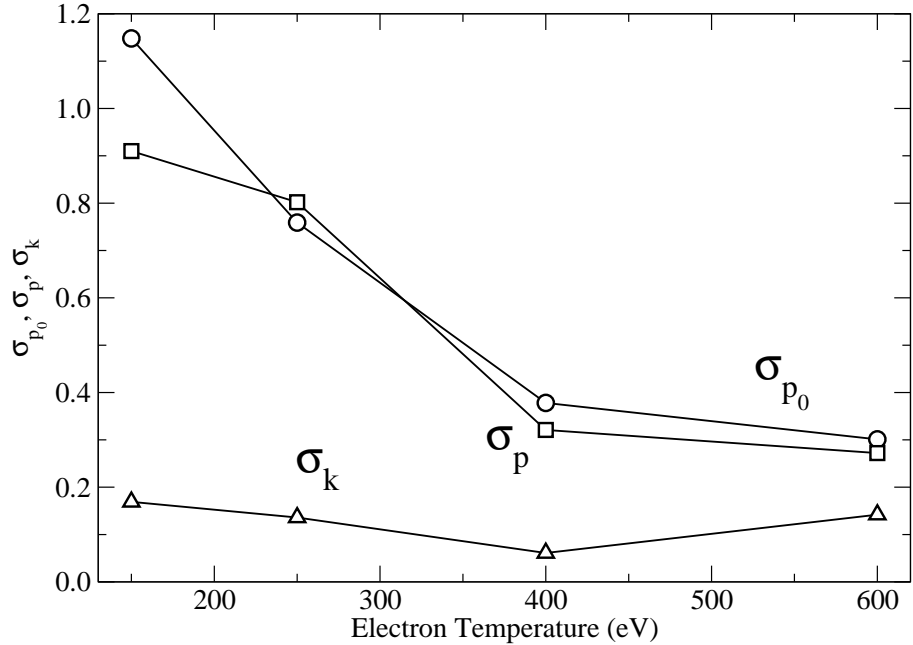}
\vspace*{1cm}
\caption{Standard deviation $\sigma$ for calculated parameters $p_0$ 
(circles), $p$ (squares), and $k$ (triangles) as a function of electron 
temperature for the germanium case at electron density 
N$_e$ = 10$^{17}$ cm$^{-3}$.}
\label{fig2}
\end{figure*}

Finally, consider the dependence of $\sigma$ on electron temperature $T_e$. The
SAHA database contains data for $T_e$ = 150 eV, 250 eV, 450 eV, and 600 eV at
$n_e$ = $10^{17}$ cm$^{-3}$. The calculated $\sigma$'s for $p_0$, $p$, and $k$
are presented in Fig. \ref{fig2} for the four temperatures. At low $T_e$, the
mostly populated ions are those with open shells, and since these cases are most
difficult to calculate, the difference between codes is the largest. With the
increase of electron temperature, the ionization stage approaches the
closed-shell Ne-like ion, and therefore agreement improves dramatically. Note
also that $\sigma_k$ is smaller than $\sigma_p$, and therefore in many cases a
simple one-parameter formula for $p$ would be sufficient for estimates of the
difference between codes.


\subsection{C}

Carbon cases in the SAHA database were calculated at a single electron density
of N$_e$ = $10^{22}$ cm$^{-3}$. At T$_e$ = 20 eV, which is the selected case
here, the mean ion charge varies between 1.8 to 3.3 for different codes. There
are no obvious outliers in this case, and therefore two-parameter fits were
successful for all codes (Table \ref{TbC}). Similar to the Ge case, the
correlation between calculated and fit values is generally good, although
one-parameter $p$'s for codes 8 and 9 show larger discrepancies. The standard
deviations for the different parameters $p$ in Table \ref{TbC} agree even better
than for the Ge case, however, they are about a factor of 2 larger and reach 60
\%. This would seem to be unexpected as carbon simulations could include only
seven ionization stages at most and are thus supposed to be simpler. The reason
for the larger $\sigma$'s is that the effects of ionization potential lowering
are much more important here due to the significantly higher density, and
different treatments of the continuum lowering noticeably contribute to the
increased spread of the $p$ and, to a lesser extent, $k$ values.

\begin{table}
\caption{Calculated and fit parameters $p$ and $k$ for the C case of T$_e$ = 20
eV and N$_e$ = $10^{22}$ cm$^{-3}$.  Superscript $f$ denotes fit values.
Subscript 0 denotes $p$'s determined from the one-parameter formulas (\ref{dQ})
and (\ref{newp}).}
\label{TbC} 
\begin{center}
\begin{tabular}{c|cc|cccc}
\hline 
Code No. & $p_0$ & $p_0^f$ & $p$ & $k$ & $p^f$ &$k^f$ \\
\hline 
1 & -0.795 & -1.003 & -1.182 &  0.411 & -1.440 &  0.693 \\
2 &  1.043 &  0.537 &  0.755 &  0.515 &  0.635 &  0.605 \\
3 &  0.629 &  0.434 &  0.443 &  0.295 &  0.435 &  0.361 \\
4 & -1.015 & -0.991 & -1.245 &  0.154 & -1.046 &  0.135 \\
5 & -0.770 & -0.899 & -1.064 &  0.291 & -1.001 &  0.305 \\
6 & -0.345 & -0.337 & -0.393 &  0.042 & -0.340 &  0.009 \\
7 & -0.209 & -0.395 & -0.467 &  0.331 & -0.471 &  0.374 \\
8 &  0.020 & -0.158 & -0.221 &  0.322 & -0.218 &  0.403 \\
9 &  0.257 & -0.374 & -0.132 &  0.547 & -0.362 &  0.851 \\
\hline 
$\sigma$ & 0.651 & 0.535 & 0.660 & 0.151 & 0.638 & 0.250 \\
\hline 
\end{tabular}
\end{center}
\end{table}

\subsection{Au}

The gold cases available in the SAHA database show very significant differences,
and for the selected case of N$_e$ = $10^{21}$ cm$^{-3}$ and T$_e$ = 750 eV, the
mean ion charge varies between 31 (code 7) and 44 (code 4). Moreover, ionization
distributions for codes 1 and 2 show double peak structure unlike other,
smoother bell-like distributions. It is therefore not surprising that the
absolute values of $p$, which are the main indicators of code disagreements, are
much larger than for the Ge and C cases discussed above. As a consequence, the
standard deviations for the differently determined $p$ and $k$ values do not
show the same level of agreement as previously. This situation simply reflects 
very significant differences between code results submitted to the 3rd NLTE Code
Comparison Workshop.

\begin{table}
\caption{Calculated and fit parameters $p$ and $k$ for the Au case of T$_e$ = 750
eV and N$_e$ = $10^{21}$ cm$^{-3}$. Superscript $f$ denotes fit values.
Subscript 0 denotes $p$'s determined from the one-parameter formulas (\ref{dQ})
and (\ref{newp}).}
\label{Tb Au}
\begin{center}
\begin{tabular}{c|cc|cccc}
\hline 
Code No. & $p_0$ & $p_0^f$ & $p$ & $k$ & $p^f$ &$k^f$ \\
\hline 
1 &  2.293 &  1.376 &  1.118 & -0.454 &  1.297 &  0.042 \\
2 & -0.534 & -0.410 &  0.418 & -0.085 & -0.434 & -0.025 \\
3 & -1.151 & -1.061 & -0.393 &  0.050 & -1.067 &  0.050 \\
4 & -1.338 & -1.178 & -0.497 &  0.061 & -1.189 &  0.052 \\
5 &  1.638 &  1.387 &  0.384 & -0.101 &  1.419 & -0.039 \\
6 & -1.086 & -1.042 & -0.422 &  0.070 & -1.052 &  0.069 \\
7 &  3.393 & -0.970 &  0.678 & -0.129 & -0.979 & -0.056 \\
8 & -1.162 & -0.302 & -0.302 & -0.028 & -0.218 &  0.403 \\
\hline 
$\sigma$ & 1.763 & 1.001 & 0.568 & 0.160 & 0.996 & 0.136 \\
\hline 
\end{tabular}
\end{center}
\end{table}

\section{\label{sec4}Conclusions}

As the complexity of plasma kinetic codes rapidly increases, their verification
and validation is becoming mandatory for establishing credibility of
computational results. To this end, a development of new techniques for code
comparisons is an urgent and important task. In the present paper we introduced
two new parameters for the characterization of discrepancies between plasma
kinetic codes. These parameters describe differences between effective
ionization and recombination rates used in the codes. Importantly, the only
physical quantities required for their calculation are the mean ion charges and
central momenta that are the most widely reported characteristics of plasma
kinetic calculations. Since the final formulas include only the simplest
algebra, this method provides very fast estimates of code differences in the
input atomic rates. The new parametrization was applied to the data from the 3rd
NLTE Code Comparison Workshop and the presented results clearly prove simplicity
and reliability of the method used. We plan to implement this method to the
analysis of the data from future NLTE workshops.

\section*{Acknowledgments}

This work was carried out while one of us (D.S.) spent a two-month working visit
at the Atomic Physics Division of the National Institute of Standards and
Technology. He would like to express his thanks for the cooperative atmosphere
and generous hospitality. Authors would also like to thank H. R. Griem and R. W.
Lee for reading the manuscript and valuable comments. This research was
supported in part by the Office of Fusion Energy Sciences of the US Department
of Energy.

\end{document}